\documentclass[hyper,a4paper]{JHEP3}
\usepackage{multirow}

\usepackage[dvips]{graphicx}
\usepackage{epsfig}
\usepackage{rotating}
\usepackage{amssymb}
\usepackage{amsmath}

\newcommand{\bea}{\begin{equation}}
\newcommand{\eea}{\end{equation}}
\newcommand{\be}{\begin{eqnarray}}
\newcommand{\ee}{\end{eqnarray}}
\newcommand{\nn}{\nonumber}

\def\hbar#1{\backslash\hspace{-2mm}#1}

\def\nn{\nonumber}

\catcode`\@=11
\def\lsim{\mathrel{\mathpalette\@versim<}}
\def\gsim{\mathrel{\mathpalette\@versim>}}
\def\@versim#1#2{\vcenter{\offinterlineskip
\ialign{$\m@th#1\hfil##\hfil$\crcr#2\crcr\sim\crcr } }}
\catcode`\@=12

\def\2tvec#1#2{
\left(
\begin{array}{c}
#1  \\
#2  \\
\end{array}
\right)}

\def\mat2#1#2#3#4{
\left(
\begin{array}{cc}
#1 & #2 \\
#3 & #4 \\
\end{array}
\right) }

\def\Mat3#1#2#3#4#5#6#7#8#9{
\left(
\begin{array}{ccc}
#1 & #2 & #3 \\
#4 & #5 & #6 \\
#7 & #8 & #9 \\
\end{array}
\right) }

\def\3tvec#1#2#3{
\left(
\begin{array}{c}
#1  \\
#2  \\
#3  \\
\end{array}
\right)}

\def\to{\rightarrow}

\def\ptmiss{\not\!\!{p_T}}

\def\hbar#1{\backslash\hspace{-2mm}#1}

\numberwithin{equation}{section}

\title{Higgs Signatures in 
Inverse Seesaw Model at the LHC}
\author{Priyotosh Bandyopadhyay$^{a}$, Eung Jin Chun$^b$, Hiroshi Okada$^b$ and Jong-Chul Park$^b$\\

 $^{a}$Department of Physics  and
    Helsinki Institute of Physics, University of Helsinki,
  FIN-00014, Helsinki, Finland\\
$^b$Korea Institute for Advanced Study,
Seoul 130-722, Korea\\
Email:  \email{priyotosh.bandyopadhyay@helsinki.fi, ejchun@kias.re.kr,
hokada@kias.re.kr, jcpark@kias.re.kr} }

\abstract{ In the inverse seesaw mechanism where the spontaneously
broken B-L symmetry induces tiny B-L violating Majorana masses for
right-handed neutrinos, non-standard Higgs signatures can arise
due to a possible Higgs doublet and singlet mixing and/or Higgs
boson decays to a left- and right-handed neutrino. This leads to a
remarkable feature of hadronically quiet di-lepton final states
which can exhibit, in particular, lepton flavour violating
signatures coming from flavour-dependent neutrino Yukawa
couplings. In this process, one lepton coming from the
right-handed decay could be soft enough can be missed by the
trigger level cuts of CMS and ATLAS for the di-lepton plus missing
energy signature. The prospects of such a signature are
investigated for 8 TeV and 14 TeV center of mass energy of the
LHC, taking  the maximum value of the allowed neutrino Yukawa
coupling and the right-handed neutrino mass of 100 GeV.  A PYTHIA
level simulation shows that the integrated luminosity of 10--20/fb
and 1.6/fb for 8 TeV is required to observe the inclusive leptonic
and lepton flavour violating signatures, respectively. For 14 TeV,
the reach is more and a larger parameter space of the inverse
seesaw model can be probed.}

\preprint{KIAS-P12026\\
{HIP-2012-13/TH}}

\begin{document}

\section{Introduction}

A Higgs-like boson around 125 GeV has been observed by both CMS
\cite{Higgsd1} and ATLAS \cite{Higgsd2} collaborations at the LHC.
The $5\sigma$ discovery reach has been obtained for the Higgs
boson to $\gamma\gamma$  mode for both  CMS and ATLAS. The Higgs
boson, if it is Standard Model (SM) like, can decay to di-leptonic
final states through $WW^*$ decay mode,  which is of our interest
in this paper.  However, the significance in such a channel is
still low: around $1.6\sigma$ for CMS \cite{Higgsd1}
 and $2.8\sigma$ for ATLAS \cite{Higgsd2}. The next question is, if such data will be
able to provide information of the Higgs boson property, whether it is exactly of the SM
 or deviates from the SM. Non-standard
Higgs structure arises in many extensions of the SM motivated by
various theoretical reasoning and/or experimental requirements.

In the inverse seesaw mechanism \cite{inverse-origin,mohapatra86}
explaining tiny neutrino mass, the seesaw sector couples to the SM
sector with a Yukawa coupling of order-one and thus can lead to
observable signatures at the LHC. A promising way to realize the
inverse seesaw mechanism is to consider a gauged $B-L$ symmetry
which is broken spontaneously at TeV scale
\cite{shaaban,Abdallah:2011ew}.  In this case, the $B-L$ Higgs
boson can mix with the SM Higgs boson which can change the SM
Higgs boson signature at the LHC.  Although the current LHC data
strongly restricts such a mixing, we will try to take a
conservative limit to maximize the mixing effect. Then the Higgs
sector of the inverse seesaw model contains two Higgs bosons, a
light and heavy one denoted by $h$ and $H$, respectively. As these
states are mixtures of the doublet and singlet Higgs states, their
production is governed by the usual gluon fusion process of the
doublet state.

On the other hand, the Higgs bosons can decay to the left- and
right- handed neutrinos (denoted by $\nu$ and $\Psi$,
respectively):
\begin{equation} h,H \to \nu \Psi,\; \Psi \Psi \end{equation}
or $H\to hh$ depending on their masses. Then, the right-handed neutrino allows
the decay modes:
\begin{equation} \Psi \to l W, \; \nu Z \end{equation}
which can lead to hadronically quiet final states from the Higgs
boson decay.\footnote{For related studies, see the recent works
\cite{dev12,cely12}.} A careful observation shows that the lepton
coming from the decay of the right-handed neutrino, $\Psi$ could
really be soft, depending on the mass difference between the
right-handed neutrino and the $W$ boson. We stress that, for some
parameter points, this soft lepton could be already missed by ATLAS
and CMS due to hard trigger level cuts for the leptons in the
study of the Higgs boson to $WW^*$ \cite{atlas7tev}. Another
remarkable feature of the inverse seesaw is the possibility of
lepton flavour violation as $\Psi$ can dominantly decay to a
specific lepton flavour, which occurs generically with
hierarchical neutrino Yukawa couplings. In this paper, we analyze
such non-standard Higgs signatures to discuss its discovery
potential at the 8 and 14 TeV LHC.

The paper is organized as follows. In Section 2, we introduce the
inverse seesaw model presenting the mass spectrum and the
interaction vertices. The explicit form of all the interaction
vertices of the model is shown in Appendix A. In Section 3, we
discuss experimental constraints on the neutrino Yukawa coupling
coming from the lepton universality as well as on the Higgs boson
masses coming from the LEP and current LHC data. In Section 4, the
benchmark points for our study are set up to calculate the
branching ratios for the allowed decay model, and the final state
phenomenology of each benchmark point is discussed.
In Section 5, we analyze the missing and lepton $p_T$
distributions and lepton multiplicity to suppress the standard
di-boson background. Then, we discuss the discovery potential of
the signal events of two lepton (plus missing $p_T$)
final states at the 8 and 14 TeV LHC.  We conclude in Section 6.

\section{Inverse seesaw model with U(1)$_{B-L}$}

A natural way to realize the inverse seesaw mechanism is to introduce a $B-L$ gauge
symmetry whose spontaneous breaking at the TeV scale generates a tiny $B-L$ breaking
Majorana mass for right-handed neutrinos.
The ``minimal" field content implementing this idea would be as follows:
\begin{equation}
\begin{tabular}{||c|c|c|c|c|c|c||c|c||}
\hline\hline ~~Particle~~ & ~~$Q$~~ & ~~$u^c, d^c $~~ & ~~$L$~~ & ~~$e^c, N$~~
 & ~~$S_1$~~ & ~~$S_2$~~ & ~~$\Phi $& ~~$\chi $~~\\
\hline
$Y_{B-L}$ & 1/3 & -1/3 & -1 & 1 & -1/2 & 1/2 & 0 & -1/2\\
\hline\hline
\end{tabular}
\end{equation}
where a
pair of fermionic $S_1$ and $S_2$ is required to cancel the
anomaly. The SM and $B-L$ Higgs bosons are denoted by $\Phi$ and
$\chi$, respectively.
The gauge invariance of the model allows the leptonic sector
Lagrangian in terms of the Weyl fermions as follows:
 \be
-{\cal L}= 
y_\ell  L\Phi e^c +y_\nu  L\Phi^cN+y_S N\chi S_1 + {\lambda_{S_1}
\over \Lambda} \chi^{\dagger2} S^2_1 + {\lambda_{S_2} \over
\Lambda} \chi^{2} S^2_2 + h.c.\,,
 \ee
where $\Phi^c\equiv\epsilon \Phi^*$ and $\Lambda$ is a cut-off
scale. Note that the mass term $S_1 S_2$ can be suppressed by
introducing, e.g., a discrete symmetry $Z_2$ under which $S_2$ is
odd and the others are even.
After the symmetry breaking, $\chi=(\chi^0+v')/\sqrt2$ and
$\Phi=(\phi^+,\phi)^T$ with $\phi=(\phi^0+v)/\sqrt2$, the neutrino sector
can be written as
\be {\cal L}_m^{\nu} = m_D\nu' N   + M_N N S_1+ \mu_{S} S_1^2
+h.c.\,, \ee where $m_D = y_{\nu} v/\sqrt{2}$, $ M_N =y_{S}
v'/\sqrt{2}$ and $\mu_S = \lambda_{S_1}v'^2/2\Lambda$. In the flavour basis
$\{\nu', N, S_1\}$, the $3\times 3$ neutrino mass
matrix of one generation takes the form:%
\be
\left(%
\begin{array}{ccc}
  0 & m_D & 0\\
  m_D & 0 & M_N \\
  0 & M_N & \mu_S\\
\end{array}%
\right).\label{Mnu} %
\ee%
From the diagonalization of this mass matrix, one can find the
light neutrino mass
\begin{equation}
m_\nu = \mu_S m_D^2/M_N^2 \,.
\end{equation}
Considering the constraint on $m_D/M_N$, which will be discussed
in the following section, one can obtain $m_\nu \approx 0.1$ eV
with the cut-off scale $\Lambda \approx 10^{15}$ GeV.


\medskip

For the consideration of collider phenomenology, one can simply take the limit of
$\mu_S=0$.
In this case, the active neutrino becomes massless and the other two heavy (Weyl)
neutrinos are degenerate to form a Dirac fermion.
The mass basis is defined by the following rotation
\be
\left(%
\begin{array}{c}
\nu'\\
  S_1\\
\end{array}%
\right)
=
\left(%
\begin{array}{cc}
  \cos\theta & \sin\theta \\
  - \sin\theta &  \cos\theta \\
\end{array}%
\right)
\left(%
\begin{array}{c}
\nu\\
N^c\\
\end{array}%
\right)\quad {\rm where}\quad \sin\theta\equiv
\frac{m_D}{\sqrt{m^2_D+M^2_N}},\label{sintheta} \ee
where $(N, N^c)$ form a Dirac fermion denoted by $\Psi$ with the mass
$m_\Psi =
\sqrt{m^2_D+M^2_N}$.

Now let us analyze the Higgs potential given by
\be
V(\Phi,\chi)=m^2_1\Phi^\dagger\Phi+m^2_2\chi^\dagger\chi+\lambda_1(\Phi^\dagger\Phi)^2+\lambda_2(\chi^\dagger\chi)^2
+\lambda_3(\Phi^\dagger\Phi)(\chi^\dagger\chi).
\ee
After spontaneous breaking, the mass matrix is given by
\be
M(\phi^0,\chi^0)
=
\left(%
\begin{array}{cc}
  2\lambda_1v^2& \lambda_3vv' \\
  \lambda_3vv' &   2\lambda_2v'^2 \\
\end{array}%
\right),
\ee
where we use the stationary conditions:
\be
\left.\frac{\partial V}{\partial \Phi}\right|_{v,v'}&=&0\rightarrow m^2_1+\lambda_1 v^2+\frac{1}{2}\lambda_3v'^2=0,\\
\left.\frac{\partial V}{\partial \chi}\right|_{v,v'}&=&0\rightarrow m^2_2+\lambda_2 v'^2+\frac{1}{2}\lambda_3v^2=0.
\ee
The neutral component of $\chi$ and $\phi$ mixes as follows:
\be
\left(%
\begin{array}{c}
\phi^0\\
\chi^0\\
\end{array}%
\right)
=
\left(%
\begin{array}{cc}
  \cos\alpha & \sin\alpha \\
  - \sin\alpha &  \cos\alpha \\
\end{array}%
\right)%
\left(%
\begin{array}{c}
h\\
H\\
\end{array}%
\right), \ee where $h$ ($H$) is the light (heavy) Higgs boson. The
mass eigenvalues and the mixing angle are given by
 \be
m^2_{h,H}&=&\lambda_1 v^2+\lambda_2v'^2\mp\sqrt{(\lambda_1v^2-\lambda_2v'^2)^2+\lambda^2_3v^2v'^2},\\
\tan2\alpha&=&\frac{\lambda_3vv'}{\lambda_2v'^2-\lambda_1v^2},
 \ee
where  $v=246$ GeV.
The most stringent bound on the value of the $B-L$ symmetry
breaking scale $v'$ comes from the LEP II data requiring
$m_{Z'}/g_{B-L} = |Y^{\chi}_{B-L}| v' > 6$ TeV~\cite{PDG2010}.
This tells us that $v' > 12$ TeV in our case. We set $v'=12$ TeV
for our analysis.

\medskip

The interacting terms from the Higgs kinetic Lagrangian are given by
\be
{\cal L}^K&=&|D_\mu\Phi|^2+|D_\mu\chi|^2\nn\\
&\supset&
\left[m^2_WW^+_\mu W^{-\mu}+\frac12 m^2_Z Z_\mu Z^{\mu} \right]
\left[1+\frac{c_\alpha h+s_\alpha H}{v}\right]^2 \\
&& +
\frac12 |Y^{\chi}_{B-L}|^2m^2_{Z'} Z'_\mu Z'^{\mu}
\left[1+\frac{-s_\alpha h+c_\alpha H}{v'}\right]^2\nn \,,
\ee
where $m_{Z}=gv/(2c_{\theta_W})$ and $m_{Z'}=|Y^\chi_{B-L}| g_{B-L} v'$.
The leptonic interaction Lagrangian is given by
 \be {\cal L}^{int.} &=&
\frac{y_{q_{u_i}}}{\sqrt2}(h\cos\alpha+H\sin\alpha)\bar q_{u_i}P_Rq_{u_i}+
\frac{y_{q_{b_i}}}{\sqrt2}(h\cos\alpha+H\sin\alpha)\bar q_{d_i}P_Rq_{d_i}\nn\\
&+&
\frac{y_{\nu_i}}{\sqrt2}(h\cos\alpha+H\sin\alpha)
\left[\cos\theta \bar\Psi_i P_L\nu_i +\sin\theta\bar\Psi_i P_L \Psi_i\right]\nn\\
&+&
\frac{y_{\ell_i}}{\sqrt2}(h\cos\alpha+H\sin\alpha)\bar\ell_{i}P_R \ell_i
+
\frac{y_{s_i}}{\sqrt2}(-h\sin\alpha+H\cos\alpha)
[-\sin\theta\bar\Psi_iP_L\nu_i+\cos\theta\bar\Psi_iP_L\Psi_i]
\nn\\
&+&
\frac{g}{\sqrt{2}}W^+_\mu
\left[\cos\theta\bar\nu_i\gamma^\mu P_L(U_{MNS})_{ij}  \ell_{j}+\sin\theta\bar\Psi_i\gamma^\mu P_L(U_{MNS})_{ij} \ell_{j}\right]\nn\\
&+& \left(\frac{g}{2\cos\theta_W}Z_{\mu} \right)
\left[\cos^2\theta\bar\nu_i\gamma^\mu
P_L\nu_i+\sin^2\theta\bar\Psi_i\gamma^\mu P_L \Psi_i
+\cos\theta\sin\theta(\bar\nu_i\gamma^\mu P_L \Psi_i
+\bar\Psi_i\gamma^\mu P_L \nu_i)
\right]\nn\\
&+&h.c.\,,
\ee
where $i=(e,\mu,\tau)$.
and $(U_{MNS})_{ij}$ is Maki-Nakagawa-Sakata matrix \cite{mns}.
All the interaction terms relevant for our analysis are summarized in Appendix A.

\section{Experimental constraints on $y_\nu$ and the Higgs masses}


The  inverse seesaw model allows large neutrino Yukawa couplings $y_\nu$.
The most stringent constraint on $y_\nu$ comes from the
electroweak precision data. From Table 8 of
Ref.~\cite{delAguila:2008pw}, one gets
 \be \label{ewpd} &&
\left|\frac{(m_D)_{e,\mu,\tau}}{M_N}\right| < (0.055,\ 0.057,\
0.079)\,. \ee
This tells us that $ (y_\nu)_i < c_i M_N/174\mbox{GeV}$ where
$c_i$ is the value for each flavour given in Eq.~(\ref{ewpd}). To
be consistent with this bound,  we will take
\begin{equation} \label{thetai}
\sin\theta_i \approx {(m_D)_i \over M_N} = 0.05
\end{equation}
for $i=e$ or $\mu$ for the collider study in the following sections.

\medskip

\begin{figure}[tbc]
\begin{center}
\includegraphics[width=0.65\linewidth]{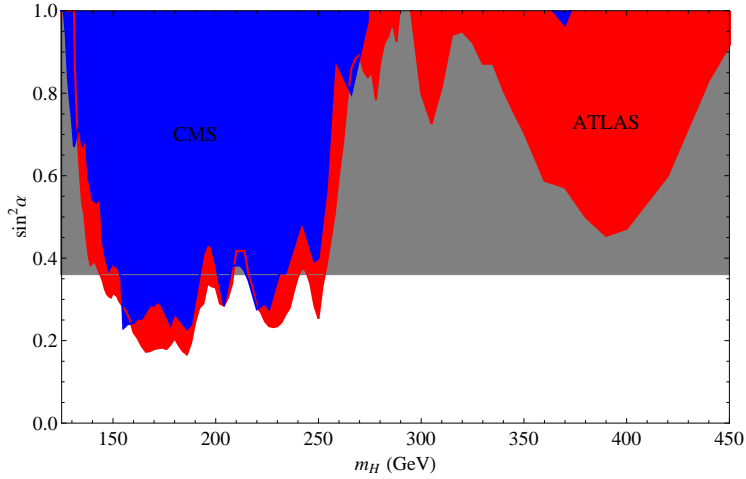}
\end{center}
\caption{The heavy 
Higgs boson mass versus $s^2_\alpha$. The red region is excluded
by ATLAS, and the blue one is excluded by CMS. The gray region is
excluded by the observation of the 125 GeV Higgs at CMS. }
\label{lhc}
\end{figure}

The extra Higgs boson mass is constrained by the LEP and LHC data
\cite{LEPb} as well as the current LHC data
\cite{Higgsd2,CMS1202}.
 In the inverse seesaw model,  the Higgs
production cross sections are suppressed by a factor of
$c_\alpha^2$ and $s_\alpha^2$ for $h$ and $H$, respectively. In
addition, the decay branching fraction into SM particles is
suppressed by a factor of $1 - {\rm Br}(h/H \rightarrow
\mbox{non-SM})$. If there exists a Higgs boson lighter than the
125 GeV Higgs, it is more singlet-like and an upper bound on
$c^2_\alpha$ is put by the LEP data \cite{LEPb}.  For the Higgs
boson mass of 50 GeV, we find $c^2_\alpha \lesssim 0.05$. For a
Higgs boson heavier than 125 GeV, the current LHC data
\cite{Higgsd2,CMS1202} puts an upper bound on $s^2_\alpha$ as
shown in Fig.~\ref{lhc}. For the 125 GeV Higgs boson, the overall
signal strength measuring the deviation from the SM prediction is
found to be $\mu = 1.4\pm 0.3$ (ATLAS) \cite{Higgsd2} and $\mu
=0.87\pm 0.23$ (CMS) \cite{Higgsd1}. Thus a non-standard Higgs
boson contribution to the 125 GeV Higgs is strongly disfavoured.
For our analysis, we will put a very conservative bound: $\mu >
0.64$, or $s^2_\alpha <0.36$ to see the maximized mixing effect.

\begin{figure}
\begin{center}
\includegraphics[width=0.55\linewidth]{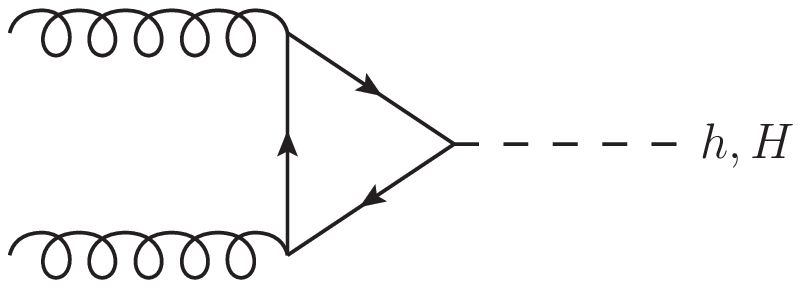}
\includegraphics[width=0.45\linewidth]{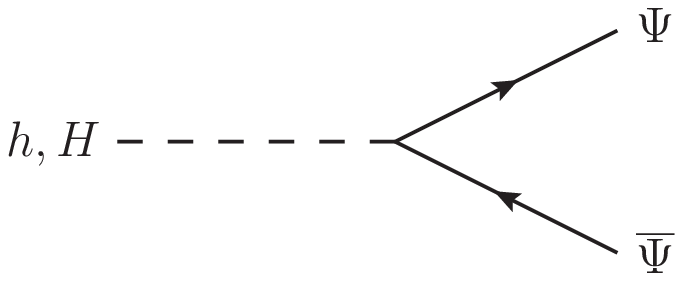}
\includegraphics[width=0.53\linewidth]{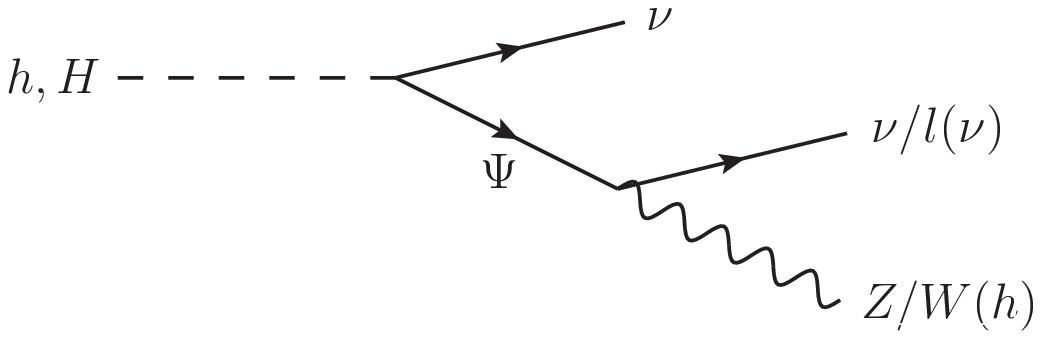}
\includegraphics[width=0.4\linewidth]{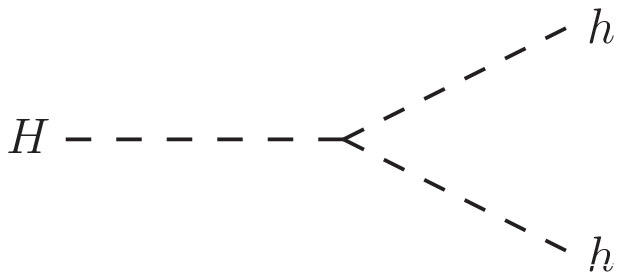}
\end{center}
\caption{Feynman diagrams of the Higgs production via gluon fusion
and the decay of Higgses.}\label{Feyndia}
\end{figure}

\begin{figure}
\begin{center}
\hskip -15pt

{\epsfig{file=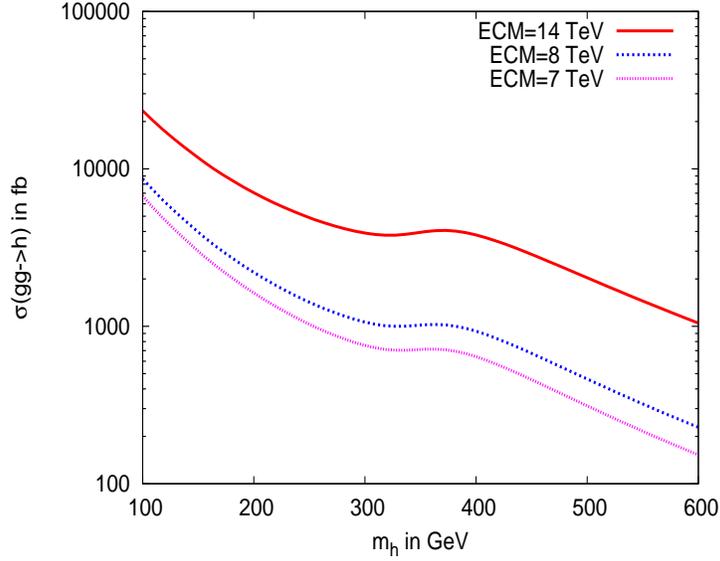,width=8 cm,height=10.0cm,angle=-90}}
\caption{Standard Model Higgs production via gluon fusion at large
Hadron collider for ECM=7, 8, 14 TeV. CTEQ5L PDF and
$Q=\sqrt{\hat{S}}$ have been used. For our Higgs boson, we have to
multiply by the corresponding rescaling factor $\sin^2{\alpha}$ or
$\cos^2{\alpha}$. }\label{Higgsprod}
\end{center}
\end{figure}

\begin{figure}
\begin{center}
\hskip -15pt {\epsfig{file=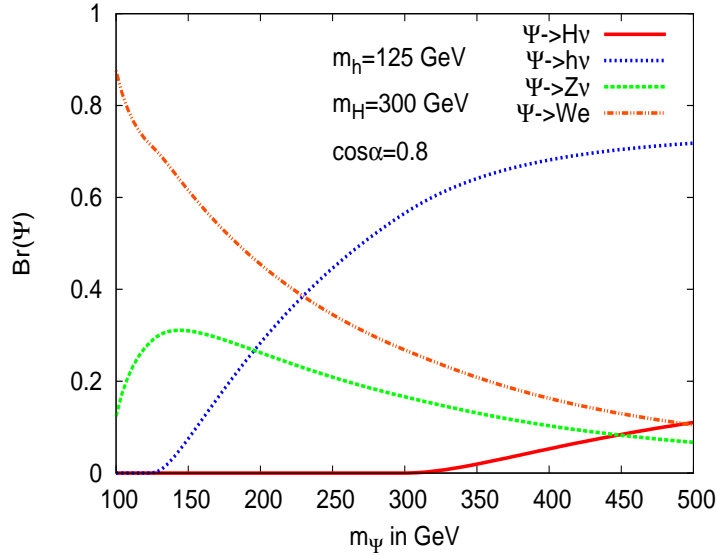,width=8
cm,height=10.0cm,angle=-90}} \caption{Variation of $\Psi$ decay
branching fraction with $m_{\Psi}$.}\label{psidbr}
\end{center}
\end{figure}

\section{Benchmark Points and final state phenomenology}

In this article, we are focusing on the gluon fusion process
leading to a light/heavy Higgs boson and its further decay to the
right-handed neutrino and the corresponding phenomenology as
stated in the introduction. Fig.~\ref{Feyndia} shows the Feynman
diagram responsible for this process. Fig.~\ref{Higgsprod}
describes the variation of the Higgs production cross-section
depending on the Higgs boson mass for 14 TeV and 8 TeV center of
mass energy (ECM) at the LHC (and also for ECM=7 TeV as a
reference). In our case, the cross-section for the
light (heavy) Higgs boson will be scaled by $\cos^2{\alpha}$
($\sin^2{\alpha}$) compared to the total Higgs production
cross-section which is shown in  Fig.~\ref{Higgsprod}. Fig.~\ref{psidbr} shows the variation of the
decay branching fraction of the right-handed neutrino, $\Psi$, with
its mass for the fixed values of Higgs masses ($m_h=125$ GeV,
$m_H=300$ GeV), $\cos{\alpha}=0.8$ and $\lambda$ couplings chosen
for the case of BP4, which is defined later. We can read from the plot
that the for low mass of the right-handed neutrino, it mostly decays to gauge bosons but for
$m_{\Psi}\geq 250$ the decay branching to $h\nu$ dominates (always
$>50\%$).

Let us now select some relevant points for the phenomenological
studies. For this purpose, we  choose $\sin\theta\approx m_D/M_N =
0.05$ close to the current bound (\ref{ewpd}) for each generation.
With this choice, we further consider two different options for the
light Higgs boson.

\begin{enumerate}
\item \underline{Very low mass light Higgs ($h$):} A very light
(non-standard) Higgs boson mass (below 114 GeV) can still satisfy
the corresponding LEP bound as discussed in the previous section.
Keeping the the light Higgs mass fixed at 50 GeV we now choose two
sets of points:

$$(m_H,m_\Psi,c_\alpha)=(125~{\rm GeV}, 100~{\rm GeV}, 0.1/0.25).$$

Thus, for these two points, it is the heavier Higgs which stays in
the discovery region of the LHC.

\item \underline{Light Higgs ($h$) in the LHC discovery region:}
Here we take $m_h=125$ GeV and choose the following combinations:

$$(m_H,m_\Psi,c_\alpha)=(200/300~{\rm GeV}, 100~{\rm GeV}, 0.8).$$

Note that the LHC bound (Fig.~\ref{lhc}) sets a limit of
$s^2_\alpha \lesssim 0.33 (0.36) $ for $m_H=200 (300) $ GeV. So,
we made a generous choice of $c_\alpha=0.8$ corresponding to
$s^2_\alpha=0.36$.

\end{enumerate}

Given the Higgs masses and  $(v,v')=(246,12000)$ GeV, the four-point Higgs
couplings for each benchmark point in Table \ref{bps} can be found as follows:
\begin{enumerate}
\item BP1: $\lambda_1$ = 0.128,\ $\lambda_2$ = $9.14 \times
    10^{-6}$,\ $\lambda_3$ = $4.42 \times 10^{-4}$,
\item BP2: $\lambda_1$ = 0.122,\  $\lambda_2$ = $1.15 \times
    10^{-5}$,\ $\lambda_3$ = $1.08 \times 10^{-3}$,
\item BP3: $\lambda_1$ = 0.202,\  $\lambda_2$ = $1.08 \times
    10^{-4}$,\ $\lambda_3$ = $3.96 \times 10^{-3}$,
\item BP4: $\lambda_1$ = 0.350,\  $\lambda_2$ = $2.20
    \times10^{-4}$,\ $\lambda_3$ =$ 1.21 \times 10^{-2}$.
\end{enumerate}
\begin{table}
\begin{center}
\renewcommand{\arraystretch}{1.4}
\begin{tabular}{||c|c|c|c|c||}
\hline\hline
Benchmark&$m_h$&$m_H$&$m_{\Psi}$&$\cos{\alpha}$\\
Points  &(GeV)&(GeV)&(GeV)&\\
\hline\hline
BP1 &50 &125 & 100 &0.1\\
\hline
BP2 &50 &125 & 100 &0.25\\
\hline
BP3 &125 &200 & 100 &0.8\\
\hline
BP4 &125 &300 & 100 &0.8\\
\hline
\hline
\end{tabular}
\caption{Benchmark points for common value of $\sin{\theta}=0.05$.}\label{bps}
\end{center}
\end{table}
Table \ref{bps} summarizes the benchmark points for our
analysis. From Table~\ref{bps}, we can see that BP1 and BP2
correspond to low mass of the light Higgs that is not excluded by
LEP~\cite{LEPb} and the heavier Higgs mass is in the discovery
region of the LHC. BP3 and BP4 correspond to relatively heavier light
Higgs $m_h=125$ GeV and heavy Higgses are 200 and 300 GeV,
respectively.

In Table~\ref{Hbr}, the decay branching fractions of the heavy
Higgs boson $H$ are shown for the benchmark points. For BP1 and
BP2, the heavy (125 GeV) Higgs decays to $\Psi\Psi$ is not open because of
the unavailable phase space, but the decays to the light Higgs boson ($h$) pair
is open. In case of BP3 where $m_H=200$ GeV, the $H$ decays to
gauge bosons via one off-shell gauge boson as before, but the decay
modes to $\Psi\Psi$ and $hh$ are closed. In case of BP4, the heavy Higgs
boson can decay to $\Psi\Psi$, as well as to $hh$.
Table \ref{hbr} shows the decay branching fractions for the light
Higgs boson $h$ for all the benchmark points. While the $h$ decay
to $\nu\Psi$ is closed in case of BP1 and BP2, it is open for
BP3 and BP4 with a branching fraction $\sim7\%$. Table~\ref{psibr} gives
the decay branching fraction of $\Psi$ for the four benchmark points. From the
Table~\ref{psibr}, it is clear that the right-handed neutrino mostly decays
to gauge bosons.

\begin{table}
\begin{center}
\renewcommand{\arraystretch}{1.4}
\begin{tabular}{||c|c|c|c|c||}
\hline\hline
Decay Modes &\multicolumn{4}{|c|}{Branching Fraction} \\
\hline
&BP1&BP2&BP3&BP4\\
\hline\hline
$\Psi\Psi$&-&-&-&0.002\\
\hline
$\nu\Psi$&0.06&0.03&0.0012&$9.1\times10^{-4}$\\
\hline
$b\bar{b}$&0.57&0.27&0.004&0.002\\
\hline
$\tau\bar{\tau}$&0.06&0.03&$4.0\times10^{-4}$&$2.1\times10^{-4}$\\
\hline
$hh$&0.035&0.55&-&0.63\\
\hline
$WW/WW^*$&0.19&0.09&0.629&0.20\\
\hline
$ZZ/ZZ^*$&0.017&0.009&0.364&0.16\\
\hline
gg &0.046&0.021&0.001&0.002\\
\hline
$\gamma\gamma$&0.002&0.001&$1.5\times10^{-4}$&$7.6\times10^{-5}$\\
\hline
\hline
\end{tabular}
\caption{Decay branching fraction of $H$ for four benchmark points.}\label{Hbr}
\end{center}
\end{table}

\begin{table}
\begin{center}
\renewcommand{\arraystretch}{1.4}
\begin{tabular}{||c|c|c|c|c||}
\hline\hline
Decay Modes &\multicolumn{4}{|c|}{Branching Fraction} \\
\hline
&BP1&BP2&BP3&BP4\\
\hline\hline
$\nu\Psi$&-&-&0.073&0.073\\
\hline
$b\bar{b}$&0.875&0.8756&0.58&0.58\\
\hline
$\tau\bar{\tau}$&0.076&0.076&0.06&0.06\\
\hline
$WW$&- &- &0.19 &0.19 \\
\hline
$ZZ$&- &- &0.017&0.017\\
\hline
gg &0.011&0.011&0.047&0.047\\
\hline
$\gamma\gamma$&0.003&0.003&0.002&0.002\\
\hline
\hline
\end{tabular}
\caption{Decay branching fraction of $h$ for four benchmark points.}\label{hbr}
\end{center}
\end{table}

\begin{table}
\begin{center}
\renewcommand{\arraystretch}{1.4}
\begin{tabular}{||c|c|c|c|c||}
\hline\hline
Decay Modes &\multicolumn{4}{|c|}{Branching Fraction} \\
\hline
&BP1&BP2&BP3&BP4\\
\hline\hline
$h\nu$&0.068&0.167&-&-\\
\hline
$Z\nu_e$&0.116&0.103&0.124&0.124\\
\hline
$W^+e$&0.817&0.730&0.876&0.876\\
\hline
\hline
\end{tabular}
\caption{Decay branching fraction of $\Psi$ for four benchmark points.}\label{psibr}
\end{center}
\end{table}

In Table~\ref{crbps}, we present the cross-sections of the Higgs
boson production for the benchmark points with the center of mass
energy of 14 and 8 TeV at the LHC, where we used  CTEQ5L as PDF and
$\sqrt{\hat{S}}$ as a scale.

\begin{table}
\begin{center}
\renewcommand{\arraystretch}{1.4}
\begin{tabular}{||c|c|c||c|c||}
\hline\hline
   &\multicolumn{2}{|c|}{ECM=14 TeV}& \multicolumn{2}{|c|}{ECM=8 TeV}\\
\hline
 & $\sigma_h $ (fb) & $\sigma_H$ (fb) & $\sigma_h$ (fb) & $\sigma_H$ (fb) \\ 
\hline\hline
BP1 & 872.33 & 15869.3& 370.18 & 5601.02\\ 
\hline
BP2 & 5452.06 & 15027.75& 2313.63 & 5304.0\\ 
\hline
BP3 &10258.94 & 2515.64& 3620.86 & 782.89\\ 
\hline
BP4 &10258.94 & 1386.12& 3620.86 & 376.92 \\
\hline
\hline
\end{tabular}
\caption{$gg\to h/H$ production cross-section for four benchmark
points for ECM=14 and 8 TeV, respectively, using CTEQ5L as PDF and
$\sqrt{\hat{S}}$ as scale.}\label{crbps}
\end{center}
\end{table}


Let us now we discuss the various final sates for  further
phenomenological studies of the benchmark points. We focus on the
contribution of the right-handed neutrino from the decays of heavy and
light Higgses depending on the benchmark points. The final states are then
determined from the decays of the right-handed neutrino.

\subsection{BP1 \& BP2}

These two benchmark points correspond to the case of $m_h=50$ GeV,
forbidding  the decay $h\to \nu \Psi$. The main decay modes of the
lighter Higgs boson are $h\to bb$ and $\tau\tau$ leading to the
final states: 2-$b$-jets + no $\ptmiss$ and 2-$\tau$-jets + no
$\ptmiss$. As the case with the usual Higgs search, it is very
difficult to isolate the former signal from the QCD backgrounds.
On the other hand, the $\tau$-jet analysis could be interesting. As
the lighter Higgs is very light in our case, it leads to very soft
jets and thus the $\tau$-jet tagging efficiency goes down.
However, $\tau$-jet tagging through one prong decay of $\tau$ can
have better handle over the QCD background. Thus, a possibility
of  {\it 2-$\tau$-jets + no $\ptmiss$} final state can still  be
important with a large accumulated luminosity.

\begin{figure}
\begin{center}
\includegraphics[width=0.75\linewidth]{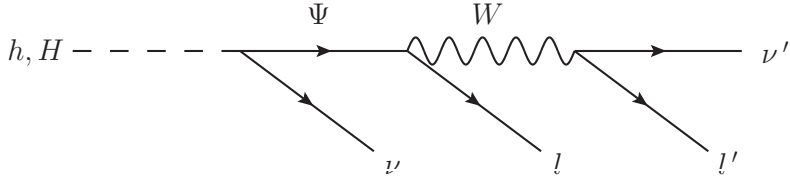}
\end{center}
\caption{Decay topology of the Higgs(es) leading to the opposite
sign lepton pair in the final state.}\label{dcchain}
\end{figure}

The non-standard Higgs signature of our interest comes from the heavy
Higgs which is SM-like with $m_{H}=125$ GeV. As can be seen from
Table~\ref{Hbr} with $m_{\Psi}=100$ GeV, the decay $H\to \nu \Psi$
has a  branching fraction $\sim 6\%$ for BP1 and
$~3\%$ for BP2. Then, the produced right-handed neutrino ($\Psi$)
will decay to gauge bosons as we can read from Table \ref{psibr}.
This will lead to the prominent signal with respect to the
backgrounds (see Fig.~\ref{dcchain}):
 \bea\label{sig1} 2\ell\, \,+\,\mbox{0-jet} \,+
\,\ptmiss. \eea
Hadronically quiet criteria along with two leptons kills the dangerous QCD
and other hadronic backgrounds. The major SM hadronically quiet
backgrounds come from the leptonic decays of $WW$, $WZ$ and $ZZ$.
Although there is a  huge single  vector boson production leading
to two leptons; $pp\to \gamma^*/Z\to \ell^+\ell^-$, this does not
carry any missing energy and can be efficiently reducible,
specially with the $Z$ mass window cut as will be discussed later.

Let us remark that, in the case of BP2 with $\cos{\alpha}=0.25$,
the heavy Higgs ($H$) decay branching fraction to light Higgs
($h$) pair  is about 55\%. This will generate a $4b$ final state
with no $\ptmiss$. Plotting the invariant mass distribution of
$b$-jet pairs, one could find a $h$ peak although the signals
would be challenged by the QCD background.

\subsection{BP3 \& BP4}

For both BP3 and BP4, the light Higgs
boson ($h$) can decay to $\nu\Psi$ with branching fraction
$7.3\%$. This leads to a hadronically quiet di-leptonic final
state (\ref{sig1}), which was the case with the heavy Higgs for
BP1 and BP2. The hadronically quiet $2\ell$ scenario thus a
generic signature for all the benchmark points.

On the other hand, the heavy Higgs ($H$) decay branching fraction to
$\nu\Psi$ drops to $< 1\%$. In case of BP4, the heavy Higgs decay
to $\Psi\Psi$ opens up but with very small branching fraction
$\sim 0.2\%$. Even with such small branching fractions, the
$\Psi\Psi$ and $\nu\Psi$ modes can still be probable due to the
large production cross-section as in Table \ref{crbps}.  BP4
allows the heavier Higgs decay mode $H\to hh$ with $63\%$ of decay
branching fraction, which is closed for BP3. The subsequent light Higgs decay to $\nu \Psi$ will  open up
an additional final state:
  \bea\label{sig2} 4\ell\, \,+\,\mbox{0-jet} \,+
\,\ptmiss. \eea
The $4\ell$ final state in case of benchmark points 3 and 4
also comes from the heavier Higgs decay to gauge boson pair.
This implies that for heavier Higgs scenarios, either through
gauge boson pairs or through the right-handed neutrino pair,
$2\ell/4\ell$ final states without any hadronic activity are generic
ones. In the following section, we will focus on these
hadronically quiet two lepton final states for a collider simulation
as the four lepton states turn out to be less significant.

\section{Results}

In this section, we go through a PYTHIA level analysis before
presenting the final state results. {\tt PYTHIA (version 6.4.22)
\cite{PYTHIA}} has been used for the purpose of event generation.
The model is implemented in Calchep \cite{calchep} and the corresponding
mass spectrum and decay branching fractions are generated. These are
then fed to {\tt PYTHIA} by using the SLHA interface \cite{Skands-SLHA}.
Subsequent decays of the produced particles, hadronization
and the collider analysis were performed using {\tt PYTHIA}.
We used {\tt CTEQ5L} parton distribution function (PDF) \cite{CTEQ-PDF}
for the analysis. The renormalization/factorization scale $Q$ was
chosen to be the parton level center of mass energy, $\sqrt{\hat{S}}$.
We also kept ISR, FSR and multiple interaction on for the analysis.
We have used {\tt PYCELL}, the toy calorimeter simulation provided in
{\tt PYTHIA}, with the following criteria:

\noindent
I. The calorimeter coverage is $\rm |\eta| < 4.5$ and the segmentation is
given by $\rm\Delta\eta\times\Delta\phi= 0.09 \times 0.09 $ which resembles
a generic LHC detector.

\noindent
II. $\Delta R \equiv \sqrt{(\Delta\eta)^{2}+(\Delta\phi)^{2}} = 0.5$
        has been used in cone algorithm for jet finding.

\noindent
III. $p_{T,min}^{jet} = 20$ GeV.

\noindent
IV. No jet matches with a hard lepton in the event.

In addition, the following set of standard kinematic cuts
were incorporated throughout:

\noindent
1. $p_T^{\ell} \geq 5$ GeV and $\rm |\eta| _{\ell} \le 2.5$,

\noindent
2. $|\eta| _{j}\leq 2.5$, $\Delta R_{\ell j} \geq 0.4$,
$\Delta R_{\ell\ell}\geq 0.2,~$

\noindent
where $\Delta R_{\ell j}$ and $\Delta R_{\ell \ell}$ measure the lepton-jet and
lepton-lepton isolation, respectively. Events with isolated leptons, having $p_T\ge 5$ GeV, are taken for the final state analysis.

\begin{figure}
\begin{center}
{\epsfig{file=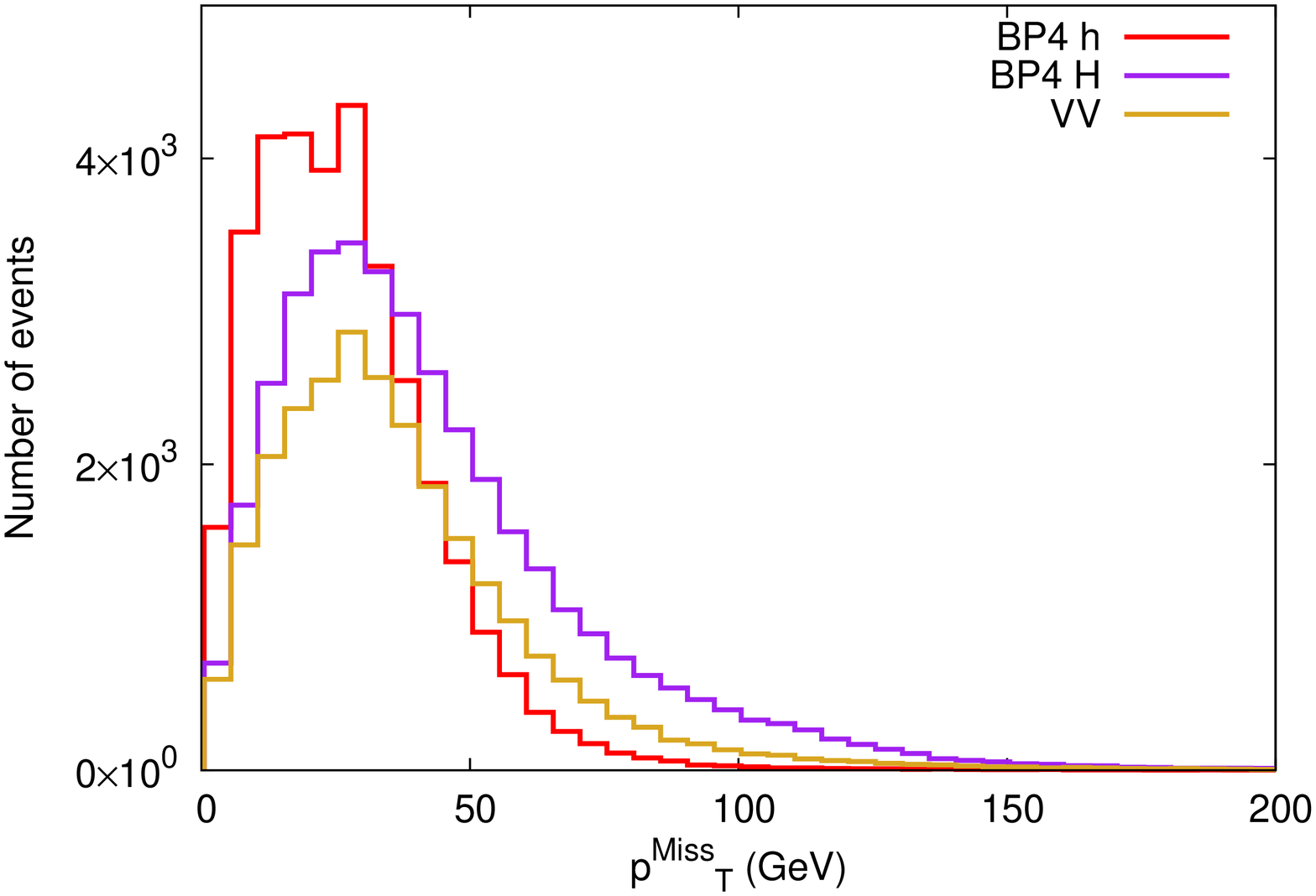,width=7.3cm,height=7.8cm,angle=-90}}
\hskip -12pt
{\epsfig{file=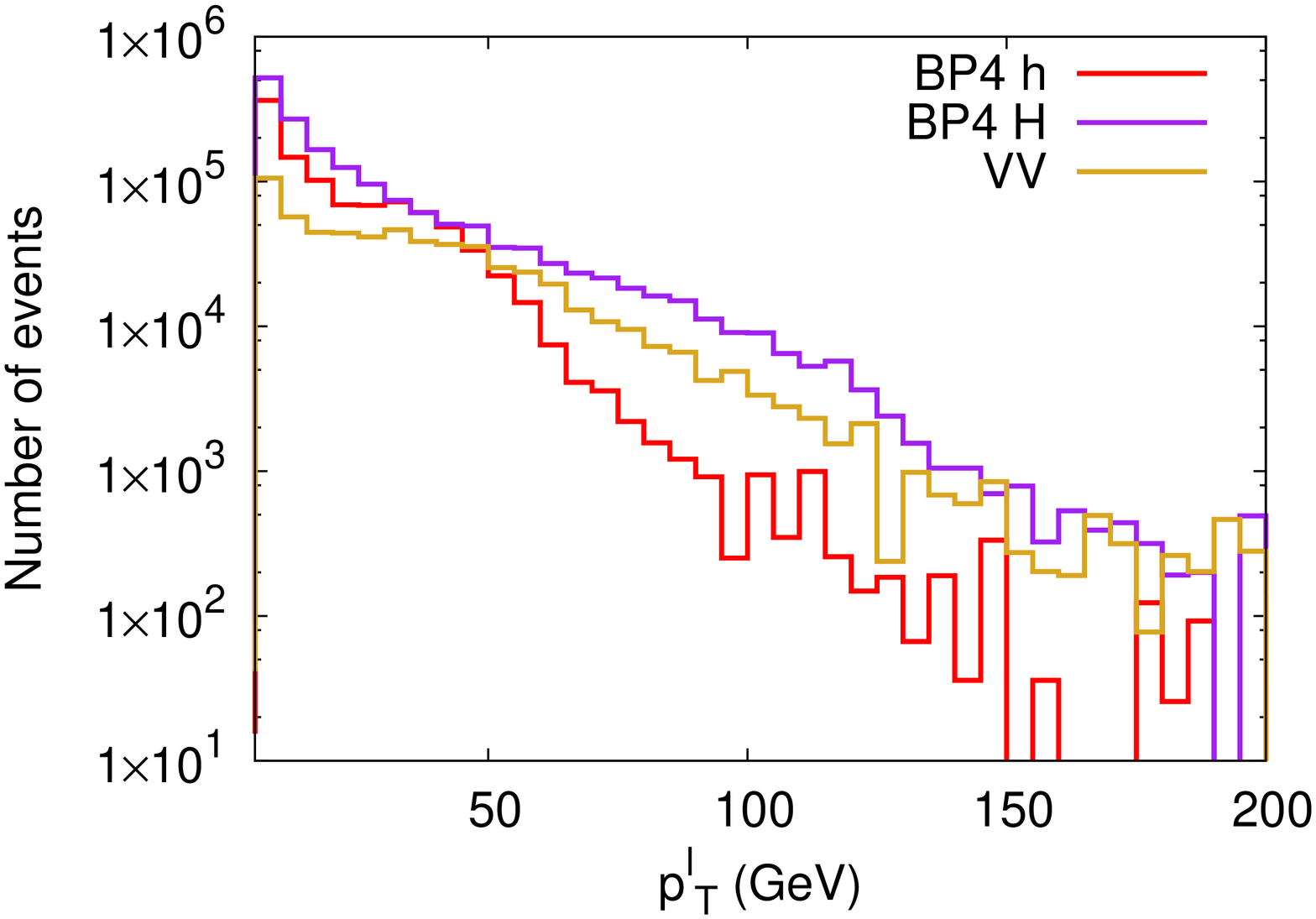,width=7.2cm,height=7.6cm,angle=-90}}
\caption{$\ptmiss$ distribution (left) from heavy ($H$) and light ($h$) Higgses for BP4 and gauge boson pair background. The lepton $p_T$ distribution  (right) of signal$\times 120$ for BP4 heavier Higgs ($H$), signal$\times 10$  for the lighter Higgs ($h$) and background at 10 fb$^{-1}$ integrated luminosity. Both of the plots are for ECM=8 TeV.}
\label{lpt}
\end{center}
\end{figure}


\begin{figure}
\begin{center}
{\epsfig{file=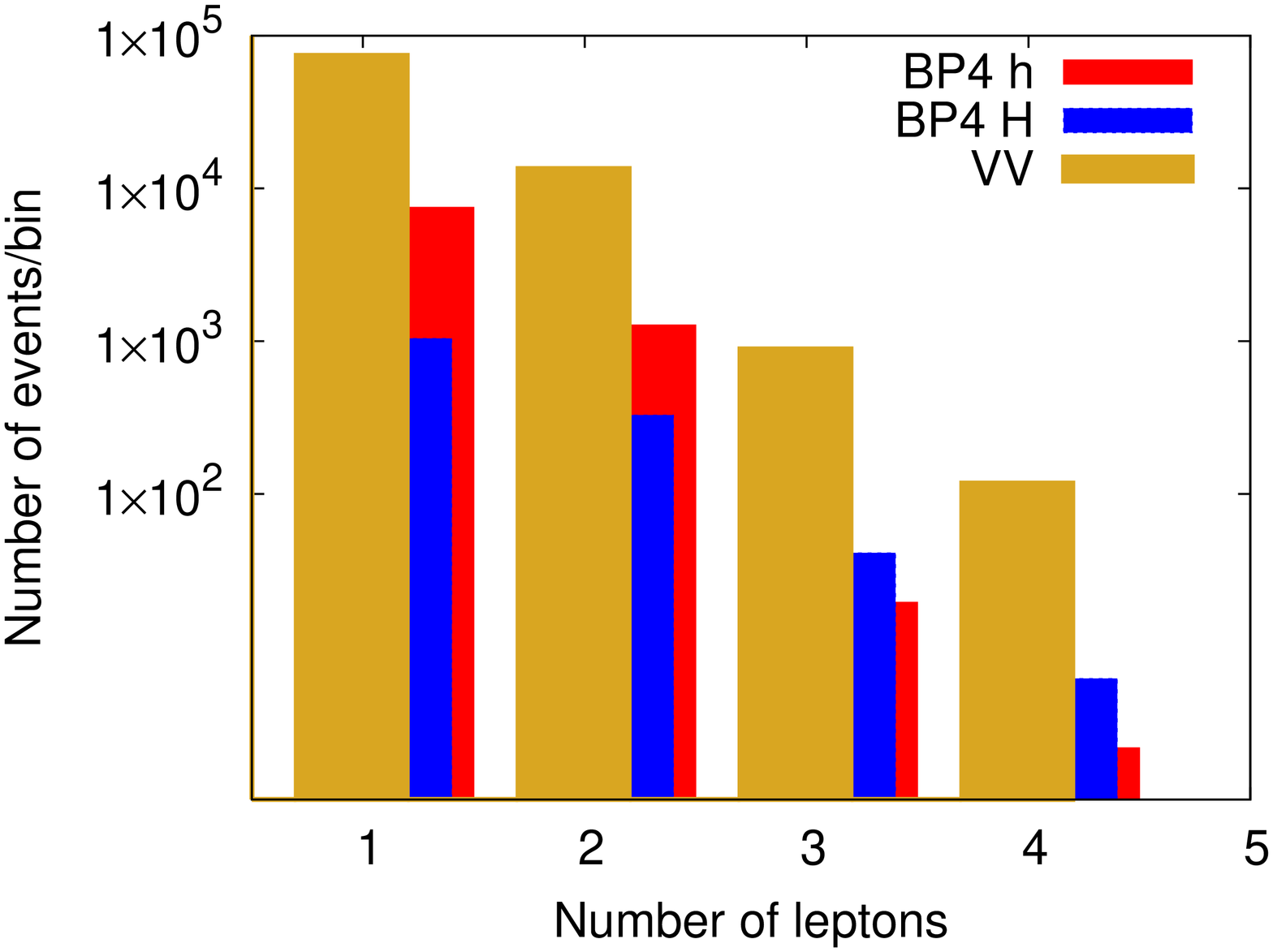,width=7.5 cm,height=7.5cm,angle=-90}}
\hskip -12pt
{\epsfig{file=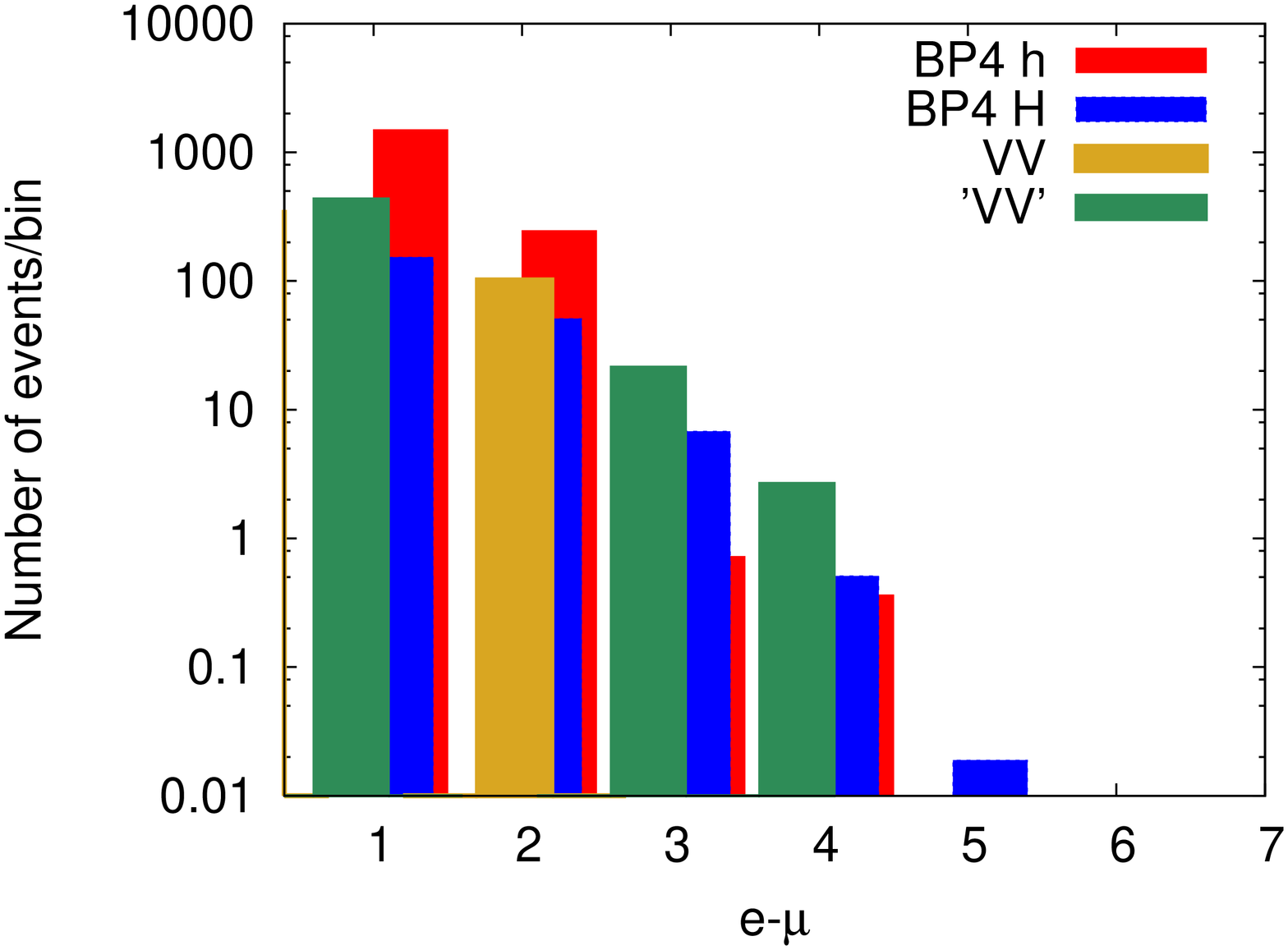,width=7.5cm,height=7.5cm,angle=-90}}
\caption{The left figure shows the lepton multiplicity distribution of signal for BP4 heavier Higgs ($H$), for the lighter Higgs ($h$)  and background at 10 fb$^{-1}$ integrated luminosity with ECM=8 TeV and the right figure shows $n_e-n_\mu$ ($n_\mu-n_e$) from the signal and di-boson background denoted by VV ('VV').  }
\label{lmc}
\end{center}
\end{figure}

Fig.~\ref{lpt} (left) plot shows the  $\ptmiss$ distribution
for the model for BP4 and the SM background which is coming from
the gauge boson pairs ($WW,ZZ,ZW$). In both cases, the origin of $\ptmiss$
is neutrinos. In the case of background, it comes from the decay
of the gauge bosons and thus it peaks around $45$ GeV. On the other hand,
in the case of signal,
there are two different sources of neutrinos. The ones from
the gauge bosons will behave similarly but the ones
from the Higgs decay ($H \to \Psi\, \nu $) can have more $p_T$,
depending on the mass difference between the Higgs and
the right-handed neutrino. This results in a tail at higher $\ptmiss$ (see Fig.~\ref{lpt} (left)) and  some enhancement in the $\ptmiss$ distribution for the heavier Higgs signal around 100 GeV for the case of benchmark point 4. In comparison, for
the lighter Higgs ($h$), the tail dies below 100 GeV as expected.
A requirement of minimum $\ptmiss$  could be a good handle to kill the leptons
coming from single $Z$, as charged lepton pair and neutrino pair come from $Z$
decay in mutually exclusive manner. Higher $\ptmiss$ cut $\geq 100$ GeV could
be a good handle to search for the existing heavy Higgs ($H$), however, we use
soft  $\ptmiss$ cut while looking for 125 GeV Higgs as discussed in the next
section.

Fig.~\ref{lpt} (right) plot shows the lepton $p_T$ distribution
coming from the signal ($H$ and $h$) in the case of BP4 and from the
gauge boson pair background, respectively. In the case of signal, the leptons
that we will be looking for, have different origin. The one coming from the right-handed neutrino can be very soft due to the small mass gap between the right-handed neutrino and the $W$ boson.  The other one coming from $W$ decay would be
of SM like.  We can see from the figure that the
lepton from the signal, specially for the case of the heavier Higgs ($H$), is
as hard as 125-150 GeV. This is because for BP4 the heavier Higgs mass is 300 GeV. The decay of such a particle to the light Higgs ($h$) pair or gauge boson pairs or though $\Psi\Psi$ or $\nu\Psi$ will share the momentum. Now the boosted
decay products will transfer their momentum to their daughter leptons, which
carry the momentum depending on the decay channels. On the other hand, the lepton coming from the light Higgs ($h$) decay could be of very small momentum due to the small mass difference between the Higgs and the right-handed neutrino, which decays
further to a lepton and $W$.  Thus, the lepton $p_T$ cuts, upper for the soft leptons
and lower for hard leptons, will be crucial  for signal event selection and reduction of the backgrounds.

Fig.~\ref{lmc} (left) describes the lepton multiplicity distribution
for the signal for BP4 coming from the heavier Higgs and the lighter
Higgs and also from the gauge boson pair backgrounds, respectively. The
lepton multiplicity distribution is independent of the relative charge,
i.e., same sign or opposite sign. We can see from the Fig.~\ref{lmc} (left)
that the gauge boson pair backgrounds also have $2l$ and $4l$ final states
from their decay to the charged leptons. Fig.~\ref{lmc} (right) describes
the electron and muon number difference. The signal events contain more electrons for
the signal from the decay of the right-handed neutrino as it is assumed to couple only to
one flavour, i.e, the electron. On the other hand, for the background we expect
the number to be very similar for $e$ and $\mu$ case (that is, $|n_e-n_\mu| \ll n_e+n_\mu$).
From Fig.~\ref{lmc} (left and right), we can see that the
background is reduced a lot, in particular, for the case of $n_e-n_\mu$.

The main Standard Model backgrounds come from the di-boson production as discussed earlier. Apart from these,
the Drell-Yan, i.e., $pp\to Z/\gamma^*\to \ell^+\ell^-$ could have
been a major background due to large cross-section but we find
that this background is reducible one as it gives di-leptons without missing energy.
We will also
have a window cut $|M_{\ell,\ell}- M_Z|> 5$ GeV to further kill
the dominant $Z$ induced  background. The $t\bar{t}$ background which
is sub-dominant but still comparable to the signal event has also been
implemented.

In the following two subsections, we present two sets
of results. First, we discuss the hadronically quiet di-lepton
scenario for inclusive flavour. Then, we address the case of
lepton flavour violation, that is,
$y_e \gg y_{\mu, \tau}$ (or $y_\mu \gg y_{e, \tau}$) for which
the right-handed neutrino will
decay only to $eW$ and $\nu_e Z$ (or to $\mu W$ and $\nu_\mu Z$).

\subsection{Inclusive charged lepton signature}

\begin{table}
\begin{center}
\renewcommand{\arraystretch}{1.2}
\begin{tabular}{||c||c|c|c|c|c||c|c|c||}
\hline\hline 8TeV/10fb$^{-1}$ &
\multicolumn{5}{|c|}{Signal}&\multicolumn{3}{|c|}{Background}\\\hline
Final state&  & BP1  & BP2 & BP3 & BP4 & VV&$Z/\gamma^*$&$t\bar{t}$ \\
\hline \hline \multirow{2}{*}{S1}& $h$ &0.09 &0.23 &269.03&269.03
&
\multirow{2}{*}{3558.47}&\multirow{2}{*}{576.89}&\multirow{2}{*}{209.29}\\
($2l$) & $H$  &352.02 &154.34 & 43.75 &4.60&&& \\ \hline
Significance &  & 5.1 & 2.3 & 4.6 & 4.0 &&& \\
\hline \hline 14TeV/10fb$^{-1}$
&\multicolumn{5}{c}{Signal}&\multicolumn{3}{|c|}{Background}\\\hline
Final state& & BP1  & BP2 & BP3 & BP4 & VV&$Z/\gamma^*$&$t\bar{t}$ \\
\hline \hline \multirow{2}{*}{S1 }& $h$ &42.88&267.97 &600.66
&600.66 &
\multirow{2}{*}{5774.59}&\multirow{2}{*}{3388.27}&\multirow{2}{*}{546.85}\\
($2l$) & $H$  &820.44 &324.60 & 117.48 &14.34&&&\\ \hline
Significance& & 8.4 & 5.8 & 7.0 & 6.0 &&&\\
\hline \hline

\end{tabular}
\caption{Number of events for $2\ell$ final states for  the
benchmark points and the SM backgrounds at an integrated
luminosity of 10 fb$^{-1}$ with ECM = 8 and 14
TeV.}\label{fsn}\label{fsn14t}
\end{center}
\end{table}

In this section, we discuss the flavour independent
results which include both electron and muon, though
we have lepton flavour violating coupling as the right-handed neutrino
couples mostly to electrons, i.e., $y_e \gg y_{\mu, \tau}$. This study is
 also suitable for the case of $y_e = y_\mu$.

For BP1 and BP2, $h\to \nu\Psi$ decay is closed
and  $h$ dominantly decays to $b\bar{b}$
or $\tau\bar{\tau}$ (see Table~\ref{hbr}). The only
contribution to $2\ell$ comes from the semi-leptonic decays of $b$
and $\tau$. Thus, we do not have large contribution from the light
Higgs in these cases. BP2 has more events than BP1 in the final state
due to the larger cross-section compared to BP1. On the other hand, the
heavier Higgs ($H$) can decay to $\nu\Psi$ with small branching
fraction ($<10\%$) as described in Table~\ref{Hbr}. Thus, the contributions
 are large enough due to large production cross-sections for the
heavier Higgs in the cases of BP1 and BP2 as can be seen from Table~\ref{crbps}.
However, in the final event counting we expect BP1 should have twice the event of BP2,
as the Br$(H\to\Psi\nu)\sim 6\%$ in the case of BP1 and which is 3\% in BP2
as cane be read from Table~\ref{Hbr}.

BP3 and BP4 are exactly same in terms of the lighter Higgs ($h$)
having $m_h=125$ GeV and $c_{\alpha}=0.8$. This leads to the same decay branching
to $\nu\Psi$, $7.3\%$,  which results in large number of
events for the final state. Also the gauge boson pairs contribute to the final
states through off-shell decays of the lighter Higgs ($h$).
In the case of the heavier Higgs ($H$), BP3 has the cross-section almost twice
as large as BP4 (see Table~\ref{crbps}), and the branching fraction
to  gauge boson pair is around $\sim 99\%$ for BP3 and $\sim 36 \%$ for BP4.
Unlike BP3, BP4 allows the decay $H\to hh$ with the branching fraction of $63\%$,
which gives additional contribution to the di-lepton final state.

As we have discussed earlier, the lepton coming from the right-handed
neutrino will be softer compared with that coming from $W^\pm$ (see Fig.~\ref{lpt}).
Thus, we demand the softer lepton with $p_T\leq 30$ GeV and the harder
lepton with $p_T\geq 20$ GeV. This would help to reduce the SM di-lepton backgrounds, which  generically come from gauge-boson decays. To kill the $Z$ boson background ,
we reject the events with lepton invariant mass ($M_{\ell,\ell}$) around $Z$
mass window, i.e., $|M_{\ell,\ell}-M_Z|\leq 5$ GeV (defined as $n_Z=0$). On top of that, we also demand $p_T^{\ell_1}+p_T^{\ell_2}\leq 100 $ GeV and
$\ptmiss \geq 30$ GeV. When we look for the signal coming from 125 GeV Higgs, we demand
the sum of lepton $p_T$ and $\ptmiss$ should be less than 125 GeV. This cut will exclusively select
events coming from 125 GeV Higgs. When we look for the events coming directly
from heavier Higgs ($H$: $m_{H}=200, 300$ GeV), the cut can be modified accordingly.
In summary, the selection cut used for the inclusive $2l$ analysis is as follows:
\bea \label{eq:S1}
\rm{S1}:\, n_{\ell}=2,\, n_{jets}=0,\,
n_{Z}=0,\, \ptmiss \geq 30 \rm{GeV},\, P_T^{\ell} \leq 100
\rm{GeV},\, M_{eff} \leq 125 \rm{GeV} \eea where $M_{\rm{eff}}=
\sum (p_T^{\ell} + \ptmiss)$ and $P_T^{\ell}=\sum p_T^{\ell}$.

In Table~\ref{fsn}, we present all the event numbers from the signal (S) and background (B), and
the significance defined by $S/\sqrt{S+B}$ for an integrated luminosity of 10 fb$^{-1}$ at the
8 and 14 TeV LHC. We can see that the behaviour of BP1 and BP2
are complementary to that of BP3 and BP4 as the main contribution to the signal events come
either from the heavy Higgs ($H$) for BP1 and BP2 or from the light Higgs  ($h$) for BP3 and BP4,
and the sum of two contributions are more or less the same except for BP2.
The signal significance at the 8 TeV LHC
reaches around $5\sigma$ except for BP2.  In the case of the 14 TeV LHC,
all the benchmark points reach high significance at an integrated luminosity of 10 fb$^{-1}$,
which implies that the parameter space of the inverse seesaw model can be readily
 probed at the LHC.

\subsection{Lepton Flavour Violating signature}

\begin{table}
\begin{center}
\renewcommand{\arraystretch}{1.2}

\begin{tabular}{||c||c|c|c|c|c||c|c|c||}
\hline\hline 8TeV/10fb$^{-1}$
&\multicolumn{5}{|c|}{Signal}&\multicolumn{3}{|c|}{Background}\\\hline
Final state&  & BP1  & BP2 & BP3 & BP4 & VV&$Z/\gamma^*$&$t\bar{t}$ \\
\hline \hline \multirow{2}{*}{S2}& $h$ &0.04&0.0 &106.82 &{106.82}
&
\multirow{2}{*}{878.24}&\multirow{2}{*}{263.3}&\multirow{2}{*}{59.03}\\
($2e$)& $H$  &122.38 &55.96 & 10.44 &1.58&&&\\ \hline
Significance & & 3.4 & 1.6 & 3.2 & 3.0 &&& \\
\hline \hline 14TeV/10fb$^{-1}$
&\multicolumn{5}{|c|}{Signal}&\multicolumn{3}{|c|}{Background}\\\hline
Final state& & BP1  & BP2 & BP3 & BP4 & VV&$Z/\gamma^*$&$t\bar{t}$ \\
\hline \hline \multirow{2}{*}{S2}& $h$ &11.00&68.70 &242.62
&242.62 &
\multirow{2}{*}{1480.36}&\multirow{2}{*}{1602.50}&\multirow{2}{*}{123.78}\\
($2e$)& $H$  &311.83 &125.48 & 26.67 &3.81&&&\\ \hline
Significance & & 5.4 & 3.3 & 4.6 & 4.2 &&& \\
\hline \hline
\end{tabular}
\caption{Number of events for $2e$ final states for  the benchmark
points and the SM backgrounds at an integrated luminosity of 10
fb$^{-1}$ with ECM=8 TeV and 14 TeV.}\label{fsn2e8}\label{fsn2e}
\end{center}
\end{table}

As mentioned earlier, much more significant search can be made if
the right-handed neutrino has hierarchical Yukawa couplings and thus allows
lepton flavour violation in its decay.
Let us first consider the case that the right-handed neutrino couples
only to electron. A
comparative study of electron and muon lepton flavour will give
a vital clue about the structure of the model as it can reduce significantly the
SM background.

\begin{table}
\begin{center}
\renewcommand{\arraystretch}{1.2}

\begin{tabular}{||c||c|c|c|c|c||c|c|c||}
\hline\hline 8TeV/10fb$^{-1}$
&\multicolumn{5}{|c|}{Signal}&\multicolumn{3}{|c|}{Background}\\\hline
Final state&  & BP1  & BP2 & BP3 & BP4 & VV&$Z/\gamma^*$&$t\bar{t}$ \\
\hline \hline \multirow{2}{*}{S3}& $h$ &0.04&0.23 &27.70 &27.70 &
\multirow{2}{*}{894.30}&\multirow{2}{*}{290.48}&\multirow{2}{*}{54.56}\\
($2\mu$)& $H$  &49.84 &22.80 & 11.57 &1.07&&&\\
\hline Significance & & 1.4 & 0.65 & 1.1 & 0.81 &&& \\
\hline \hline 14TeV/10fb$^{-1}$
&\multicolumn{5}{|c|}{Signal}&\multicolumn{3}{|c|}{Background}\\\hline
Final state& & BP1  & BP2 & BP3 & BP4 & VV&$Z/\gamma^*$&$t\bar{t}$ \\
\hline \hline \multirow{2}{*}{S3}& $h$ &11.03&68.97 &58.46 &58.46
&
\multirow{2}{*}{1486.24}&\multirow{2}{*}{1761.64}&\multirow{2}{*}{127.48}\\
($2\mu$)& $H$  &107.91 &48.84 & 32.20 &3.19&&&\\
\hline Significance & & 2.0 & 2.0 & 1.5 & 1.0 &&& \\
\hline\hline
\end{tabular}
\caption{Number of events for $2\mu$ final states for the
benchmark points and the  SM backgrounds at an integrated
luminosity of 10 fb$^{-1}$ with ECM=8 and 14
TeV.}\label{fsnmu8}\label{fsnmu}
\end{center}
\end{table}

For this study, we select each flavour state keeping all the other
cuts the same as  in the previous section (\ref{eq:S1}), and present
results of our analysis for each signal channel labeled as
\begin{equation}
\mbox{S2} : 2e\,; \quad \mbox{S3} : 2\mu \,; \quad \mbox{S4} :
1e+1\mu\,; \quad \mbox{and}\quad \mbox{S5} : 2e-2\mu\,.
\end{equation}
As expected, S5, corresponding to S2-S3, will give a novel
signature to probe the inverse seesaw mechanism. Instead of S5,
one could make an equivalent study with other lepton flavour
violating final states such as $2\cdot S2-S4$ or $2\cdot S3-S4$.

Table~\ref{fsn2e8} (\ref{fsnmu8}) shows the number of electron
(muon) events for the signal for all the benchmark points and the
backgrounds at an integrated luminosity of 10 fb$^{-1}$ with
center of mass energy of 8 and 14 TeV at the LHC. Unlike the
lepton-universal gauge boson decay, the right-handed neutrino will
produce only electron in the $lW$ mode. For BP1 and BP2, $h\to
\nu\Psi$ decay is not open and thus we do not have the extra
contribution to the electron events  in the case of the light Higgs boson ($h$).
This is clear from Table~\ref{fsn2e8} and Table~\ref{fsnmu8}
where the electron and muon event numbers are similar
as they come from the semi-leptonic decays of the decay products of the light Higgs ($h$).
On the other hand, the heavy Higgs ($H$) with $m_H= 125$ GeV, can
decay to $\nu \Psi$ and thus produces more electrons. The BP1
contribution is more than BP2 due to the larger branching fraction
to $\nu\Psi$ and the cross-section of the former.

\begin{table}
\begin{center}
\renewcommand{\arraystretch}{1.2}

\begin{tabular}{||c||c|c|c|c|c||c|c|c||}
\hline\hline 8TeV/10fb$^{-1}$
&\multicolumn{5}{|c|}{Signal}&\multicolumn{3}{|c|}{Background}\\\hline
Final state& & BP1  & BP2 & BP3 & BP4 & VV&$Z/\gamma^*$&$t\bar{t}$ \\
\hline \hline
\multirow{2}{*}{S4}& $h$ &0.02&0.00 &134.51 &134.51 &\multirow{2}{*}{1785.93}&\multirow{2}{*}{29.67}&\multirow{2}{*}{95.70}\\
($1 e  + 1 \mu $)& $H$  &179.79 &75.58 & 21.75 &1.94&&&\\
\hline Significance &  & 3.9 & 1.7 & 3.4 & 3.0 &&&\\
\hline \hline 14TeV/10fb$^{-1}$
&\multicolumn{5}{|c|}{Signal}&\multicolumn{3}{|c|}{Background}\\\hline
Final state& & BP1  & BP2 & BP3 & BP4 & VV&$Z/\gamma^*$&$t\bar{t}$ \\
\hline \multirow{2}{*}{S4}& $h$ &20.85&130.30 &299.56 &299.56 &
\multirow{2}{*}{2808.00}&\multirow{2}{*}{27.32}&\multirow{2}{*}{295.60}\\
($1 e  + 1 \mu $)& $H$  &400.70 &150.28 & 58.61 &7.35&&&\\
\hline Significance &  & 7.1 & 4.8 & 6.1 & 5.2 &&&\\
\hline \hline
\end{tabular}
\caption{Number of events for $1 e  + 1 \mu $ final states for
the benchmark points and the SM backgrounds at an integrated
luminosity of 10 fb$^{-1}$ with ECM=8 and 14
TeV.}\label{fsnem8}\label{fsnem}
\end{center}
\end{table}


Similar is the case with the lighter Higgs ($h$) contribution for
BP3 and BP4 as established in Table~\ref{fsn2e8}  and
\ref{fsnmu8}. The situation for BP3 is again a little different,
as the heavier Higgs decay $H\to hh$  is not allowed and $H\to
\nu\Psi$ branching fraction is very small $\sim 0.1\%$ (see
Table~\ref{Hbr}). This ceases the extra electron contribution over
muon for BP3. For BP4, heavier Higgs mass is 300 GeV and thus
$H\to hh$ is allowed with $63\%$ branching fraction. The lighter
Higgs thus produced will decay to the right-handed neutrino as can
be read from  Table~\ref{hbr}. Similar to the previous $2\ell$
case, the complementary behaviour between the heavy and the light
Higgs remains, i.e., for BP1 and BP2, it is the heavier Higgs
which gives the lepton flavour violating signal generating extra
electrons in the final state, whereas for BP3 and BP4, it is
mainly the lighter Higgs. From Table~\ref{fsn2e8} and \ref{fsnmu8},
we can see that the significance for S3 is much smaller compared
to S4 as expected.

For completeness, we also present the event numbers for the final
state with $1e+1\mu$, defined as S4. Table~\ref{fsnem8} shows the
event numbers for S4. The Drell-Yan background is reduced a lot due
to the demand of different flavour leptons in the final state. The
signal significance in this
 case is comparable to S2 ($2e$) and
greater than S3 ($2\mu$).

\begin{table}
\begin{center}
\renewcommand{\arraystretch}{1.2}

\begin{tabular}{||c||c|c|c|c|c||c|c|c||}
\hline \hline 8TeV/10fb$^{-1}$
&\multicolumn{5}{|c|}{Signal}&\multicolumn{3}{|c|}{Background}\\\hline
Final state& & BP1  & BP2 & BP3 & BP4 & VV&$Z/\gamma^*$&$t\bar{t}$ \\
\hline \hline \multirow{2}{*}{S5}& $h$ &0.00&-0.23 &79.12 &79.12 &
\multirow{2}{*}{-16.07}&\multirow{2}{*}{-27.19}&\multirow{2}{*}{4.47}\\
($2e -2\mu$)& $H$  &72.53 &33.15 & -1.13 &0.51&&&\\
\hline Significance& & 12.5 & - & 12.5 & 12.5 &&&\\
\hline \hline 14TeV/10fb$^{-1}$
&\multicolumn{5}{|c|}{Signal}&\multicolumn{3}{|c|}{Background}\\\hline
Final state& & BP1  & BP2 & BP3 & BP4 & VV&$Z/\gamma^*$&$t\bar{t}$ \\
\hline \hline \multirow{2}{*}{S5}& $h$ &-4.36&-0.27 &184.15
&184.15 &
\multirow{2}{*}{-5.87}&\multirow{2}{*}{-159.00}&\multirow{2}{*}{-3.69}\\
($2e -2\mu$)& $H$  &203.92 &76.64 & -5.53 &0.62&&&\\
\hline Significance& & 35.8 & - & 56.3 & 47.4 &&&\\
\hline \hline
\end{tabular}
\caption{Number of events for $2e -2\mu$ final states for  the
benchmark points and the SM backgrounds at an integrated
luminosity of 10 fb$^{-1}$ with ECM=8 and 14
TeV.}\label{fsnemm8}\label{fsnemm}
\end{center}
\end{table}


S2 ($2e$) from Table~\ref{fsn2e8} and S3 ($2\mu$) from
Table~\ref{fsnmu8} lead us to check the difference in the event
numbers for the $2e$ and $2\mu$ final states. As the SM
backgrounds come from the flavour-blind decays of the gauge bosons,
we expect that $2e -2 \mu$ event number will kill the background
substantially giving only the extra electrons coming from the
right-handed neutrino decays. This is the artifact of the lepton
flavour violating right-handed neutrino coupling: $y_e \gg y_{\mu,
\tau}$. Table~\ref{fsnemm8} shows the difference in number of $2e$
and $2\mu$ events (S5). We can again see that 125 GeV Higgs, which
is the heavy Higgs ($H$) for BP1 and BP2 and the light Higgs ($h$)
for BP3 and BP4, decays into more electrons than muons. The
negative sign shows the events with more muons than electrons.
From Table~\ref{fsnemm8}, we can see that $2e-2\mu$ signal has a
high significance except for BP2 for which the signal events are
overshadowed by background for both 8TeV and 14TeV LHC at an integrated
luminosity of 10 fb$^{-1}$.
\medskip

Based on the above results, we can get the results for the
opposite case where the right-handed neutrino couples only to the
muon flavour: $y_{\mu}\gg y_{e,\tau}$.  In this case, the signal
significance drops due to the muon excess in the background:
$~7\sigma$ ($10\sigma$) for all points except BP2 with the 8 (14)
TeV LHC which is still better than the inclusive flavour searches.

\subsection{Mass measurement}

\begin{figure}
\begin{center}
{\epsfig{file=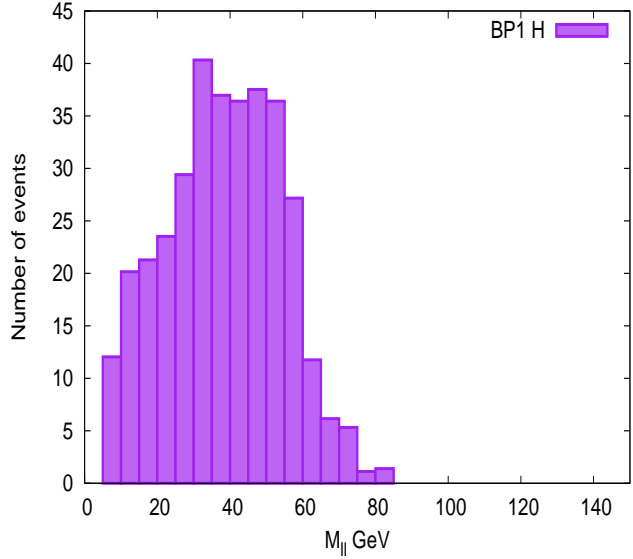,width=8.0 cm,height=9.0cm,angle=-90}}
\caption{The opposite sign lepton invariant mass  distribution
with events with the same cut as S11 for BP1 heavier Higgs ($H$)
at 10 fb$^{-1}$ integrated luminosity with ECM=8 TeV.}
\label{lepinv}
\end{center}
\end{figure}


Let us make a comment on  the prospect of measuring the
right-handed neutrino mass through the di-leptonic edge. Since we
are looking for the decay chain of the right-handed neutrino;
$\Psi \to l W \to l \bar{l}' \nu'$, the invariant mass of the
final leptons gives rise to the famous di-lepton edge \cite{MT2}:
\begin{equation}\label{mlleq}
m^{\rm{max}}_{\ell \, \ell}=m_\Psi\sqrt{1\,-\,\frac{m^2_W}{m^2_\Psi}}\sqrt{1-\frac{m^2_\nu}{m^2_W}}.
\end{equation}
As $m_\nu \approx 0$, the measurement of the maximum di-lepton
invariant mass will tell us about the value of $m_\Psi$.

Fig.~\ref{lepinv} shows the distribution of the di-lepton
invariant mass for opposite sign di-leptons from the signals only
under the same cuts as used for S1. Clearly, the edge can be seen
at 60 GeV. Now reconstructing  $W$ through its hadron decay mode
would be crucial in detecting the decay topology. For the signal,
the decay topology via $W$ boson indicates the mass of the
right-handed neutrino at 100 GeV from the Eq.~(\ref{mlleq}). Thus,
the prescription described above not only gives a variable to get
signal events over backgrounds but also results in a possibility
to measure the mass of the right-handed neutrino. Analyzing the
signal and background for events at 10 fb$^{-1}$ of integrated
luminosity, however, we find that it is very difficult to recognize the
di-leptonic edge. We have to go to much higher luminosity to get a
clear edge from the total distribution.

\section{Conclusion}

The inverse seesaw model introduces a tiny $B-L$ breaking Majorana
mass of the right-handed neutrino which explains the smallness of
the neutrino mass. This allows rather large neutrino Yukawa
couplings through which the right-handed neutrinos  can be
produced at the LHC.  We point out that the di-lepton final state
with missing energy could be a smoking gun signal probing the
Higgs and the right-handed neutrino of the inverse seesaw model.
Furthermore, a novel signature of lepton flavor violation in these
final states, such as the difference in the $ee$ and $\mu\mu$
event numbers, may occur due to flavor-dependent neutrino Yukawa
couplings. Taking the neutrino Yukawa coupling $y_{e,\mu}=0.029$,
which is close to the upper limit put by the lepton universality,
and the right-handed neutrino mass of 100 GeV, we studied the LHC
prospects to look for the inverse seesaw model in the four
benchmark points. Performing a PYTHIA level simulation, it is
found that the $5\sigma$ signal significance can be achieved with
the integrated luminosity of $10-20$ fb$^{-1}$ for the inclusive
lepton (flavour-blind) signature, and $\lsim 2$ fb$^{-1}$ for the
flavour violating signature at the 8 TeV LHC. We also pointed out
that the observation of the di-lepton edge could shed light on
the right-handed neutrino mass measurement. But it turns out to be
hard to see the di-leptonic edge over the background with the
nominal luminosity of the LHC14.

\medskip

{\bf Acknowledgement}: EJC was supported by the National Research
Foundation of Korea (NRF) grant funded by the Korea government
(MEST) (No.~20120001177). PB thanks Korea Institute for Advanced Study
for the travel support and local hospitality during some parts of this work.


\appendix

\section{Vertices}

We write down all the relevant vertices below:
\be
&&h-\bar q_{u(d)_i}-q_{u(d)_i}\ :\ \frac{y_{q_{u(d)_i}}\cos\alpha }{\sqrt2}P_R,\\
&&H-\bar q_{u(d)_i}-q_{u(d)_i}\ :\ \frac{y_{q_{u(d)_i}}\sin\alpha }{\sqrt2}P_R,\\
&&h-\bar \ell_i-\ell_i\ :\ \ \frac{y_{\ell_i}\cos\alpha }{\sqrt2}P_R,\\
&&H-\bar \ell_i-\ell_i\ :\ \ \frac{y_{\ell_i}\sin\alpha }{\sqrt2}P_R,\\
&&h-\bar\Psi_i-\nu_i \ :\ \ \frac{y_{\nu_i}\cos\alpha\cos\theta +y_{s_i}\sin\alpha\sin\theta}{\sqrt2}P_L,\\
&&H-\bar\Psi_i-\nu_i \ :\ \ \frac{y_{\nu_i}\sin\alpha\cos\theta-y_{s_i}\cos\alpha\sin\theta}{\sqrt2}P_L,\\
&&h-\bar\Psi_i-\Psi_i \ :\ \ \frac{y_{\nu_i}\cos\alpha\sin\theta-y_{s_i}\sin\alpha\cos\theta}{\sqrt2}P_L,\\
&&H-\bar\Psi_i-\Psi_i \ :\ \ \frac{y_{\nu_i}\sin\alpha\sin\theta +y_{s_i}\cos\alpha\cos\theta}{\sqrt2}P_L,\\
&&W^+ -\bar \nu_i-\ell_j \ :\ \ \frac{g\cos\theta (U_{MNS})_{ij} }{\sqrt2}P_L,\\
&&W^+ -\bar \Psi_i-\ell_j \ :\ \ \frac{g\sin\theta (U_{MNS})_{ij} }{\sqrt2}P_L,\\
&&Z -\bar \nu_i-\nu_i \ :\ \ \frac{g\cos^2\theta }{2\cos\theta_W}P_L,\\
&&Z -\bar \Psi_i-\Psi_i \ :\ \ \frac{g\sin^2\theta }{2\cos\theta_W}P_L,\\
&&Z -\bar \nu_i-\Psi_i(=Z -\bar \Psi_i-\nu_i ) \ :\ \ \frac{g\cos\theta\sin\theta }{2\cos\theta_W}P_L,
\ee
\be
&&h^4 \ :\ 6[\lambda_1c^4_{\alpha}+\lambda_2s^4_{\alpha}+\lambda_3c^2_{\alpha}s^2_{\alpha}],\\
&&H^4 \ :\ 6[\lambda_1s^4_{\alpha}+\lambda_2c^4_{\alpha}+\lambda_3c^2_{\alpha}s^2_{\alpha}],\\
&&h^2H^2 \ :\ 6(\lambda_1+\lambda_2)c^2_{\alpha}s^2_{\alpha}+\lambda_3(s^4_{\alpha}+c^4_{\alpha}
-4s^2_{\alpha}c^2_{\alpha}),\\
&&h^3H \ :\ (6\lambda_1-3\lambda_3)c^3_{\alpha}s_{\alpha}+
(-6\lambda_2+3\lambda_3)c_{\alpha}c^3_{\alpha},\\
&&hH^3 \ :\ (-6\lambda_2+3\lambda_3)c^3_{\alpha}s_{\alpha}+
(6\lambda_1-3\lambda_3)c_{\alpha}c^3_{\alpha},\\
&&h^3 \ :\ 6\lambda_1vc^3_{\alpha}-6\lambda_2v's^3_{\alpha}+
3\lambda_3(vs_\alpha-v'c_\alpha)s_{\alpha}c_{\alpha},\\
&&H^3 \ :\ 6\lambda_1vs^3_{\alpha}+6\lambda_2v'c^3_{\alpha}+
3\lambda_3(vc_\alpha+v's_\alpha)s_{\alpha}c_{\alpha},\\
&&hH^2 \ :\ 6\lambda_1vc_\alpha s^2_{\alpha}-6\lambda_2v's_\alpha c^2_{\alpha}+
\lambda_3[2(-vs_\alpha+v'c_\alpha)s_{\alpha}c_{\alpha}-v's^3_\alpha+vc^3_\alpha],\\
&&h^2H \ :\ 6\lambda_1vc^2_\alpha s_{\alpha}+6\lambda_2v's^2_\alpha c_{\alpha}+
\lambda_3[-2(v's_\alpha+vc_\alpha)s_{\alpha}c_{\alpha}+v'c^3_\alpha+vs^3_\alpha],
\ee
\be
&&WWh \ :\ 2m^2_W c_\alpha/v,\\
&&WWH \ :\ 2m^2_W s_\alpha/v,\\
&&WWhh \ :\ 2m^2_W c^2_\alpha/v^2,\\
&&WWhH \ :\ 2m^2_W s_\alpha c_\alpha/v^2,\\
&&WWHH \ :\ 2m^2_W s^2_\alpha/v^2,
\ee
\be
&&ZZh \ :\ 2m^2_Z c_\alpha/v,\\
&&ZZH \ :\ 2m^2_Z s_\alpha/v,\\
&&ZZhh \ :\ 2m^2_Z c^2_\alpha/v^2,\\
&&ZZhH \ :\ 2m^2_Z s_\alpha c_\alpha/v^2,\\
&&ZZHH \ :\ 2m^2_Z s^2_\alpha/v^2,
\ee
\be
&&Z'Z'h \ :\ -2|Y^{\chi}_{B-L}|^2m^2_{Z'} \frac{s_\alpha}{v'},\\
&&Z'Z'H \ :\ 2|Y^{\chi}_{B-L}|^2m^2_{Z'} \frac{c_\alpha}{v'},\\
&&Z'Z'hh \ :\ 2|Y^{\chi}_{B-L}|^2m^2_{Z'} \frac{s^2_\alpha}{v'^2},\\
&&Z'Z'hH \ :\ -2|Y^{\chi}_{B-L}|^2m^2_{Z'} s_\alpha \frac{c_\alpha}{v'^2},\\
&&Z'Z'HH \ :\ 2|Y^{\chi}_{B-L}|^2m^2_{Z'} \frac{c^2_\alpha}{v'^2},
\ee
where $Y^\chi_{B-L}=-1/2$ in our case.

\begin{thebibliography}{99}
\bibitem{Higgsd1}
  S.~Chatrchyan {\it et al.}  [CMS Collaboration],
  Phys.\ Lett.\ B {\bf 716} (2012) 30
  [arXiv:1207.7235 [hep-ex]].
\bibitem{Higgsd2}
  G.~Aad {\it et al.}  [ATLAS Collaboration],
  Phys.\ Lett.\ B {\bf 716} (2012) 1
  [arXiv:1207.7214 [hep-ex]].

\bibitem{inverse-origin}
R. N. Mohapatra, Phys. Rev. Lett. {\bf 56}, 561 (1986).

\bibitem{mohapatra86}
R. N. Mohapatra and J. W. F. Valle, Phys.\ Rev.\ D {\bf 34}, 1642
(1986).

\bibitem{shaaban}
  S.~Khalil,
  J.\ Phys.\ G {\bf 35}, 055001 (2008)
  [arXiv:hep-ph/0611205].

\bibitem{Abdallah:2011ew}
  W.~Abdallah, A.~Awad, S.~Khalil and H.~Okada,
  Eur.\  Phys.\  J.\ C {\bf }, 72:2108 (2012)
  [arXiv:1105.1047 [hep-ph]].

\bibitem{dev12}
  P.~S.~B.~Dev, R.~Franceschini and R.~N.~Mohapatra,
  arXiv:1207.2756 [hep-ph].

\bibitem{cely12}
  C.~G.~Cely, A.~Ibarra, E.~Molinaro and S.~T.~Petcov,
  arXiv:1208.3654 [hep-ph].

\bibitem{atlas7tev}
CMS Collaboration, Phys.\ Lett.\ B {\bf 710} (2012) 91 [arXiv:1202.1489 [hep-ex]];
ATLAS Collaboration, arXiv:1206.0756, ATLAS-CONF-2012-098.


\bibitem{PDG2010}
  K.~Nakamura {\it et al.}  [Particle Data Group],
  J.\ Phys.\ G {\bf 37}, 075021 (2010).

\bibitem{mns}
  Z.~Maki, M.~Nakagawa and S.~Sakata,
  Prog.\ Theor.\ Phys.\  {\bf 28}, 870 (1962).

\bibitem{delAguila:2008pw}
  F.~del Aguila, J.~de Blas and M.~Perez-Victoria,
  Phys.\ Rev.\  D {\bf 78}, 013010 (2008)
  [arXiv:0803.4008 [hep-ph]].

\bibitem{LEPb}
 R.~Barate {\it et al.}  [LEP Working Group for Higgs boson searches and
                  ALEPH Collaboration and  and],
  Phys.\ Lett.\  B {\bf 565}, 61 (2003)
  [arXiv:hep-ex/0306033].
\bibitem{CMS1202}
  S.~Chatrchyan {\it et al.}  [CMS Collaboration],
  Phys.\ Lett.\ B {\bf 710} (2012) 26
  [arXiv:1202.1488 [hep-ex]].
%
\bibitem{PYTHIA}
 T.~Sjostrand, S.~Mrenna and P.~Z.~Skands,
   ~JHEP {\bf 0605}, 026 (2006).
%

\bibitem{calchep}
  A.~Pukhov,
  hep-ph/0412191.



\bibitem{Skands-SLHA}
P.~Z.~Skands {\it et al.},
  JHEP {\bf 0407}, 036 (2004).
%
\bibitem{CTEQ-PDF}
  H.~L.~Lai {\it et al.}  [CTEQ Collaboration],
  Eur.\ Phys.\ J.\  C {\bf 12}, 375 (2000);
  J.~Pumplin {\it et al.},
  JHEP {\bf 0207}, 012 (2002).
%
\bibitem{MT2}
  C.~G.~Lester and D.~J.~Summers,
  Phys.\ Lett.\ B {\bf 463} (1999) 99
  [hep-ph/9906349].




\end{thebibliography}
\end{document}